\documentclass[useAMS,usenatbib]{mn2e}
\usepackage{times,aas_macros}
\usepackage{rotating}
\input{epsf}

\title[Modelling CBS 126]
{Analysing the \emph{Suzaku} Spectra of the Broad-Line Seyfert 1 Galaxy CBS 126}

\author[C.-Y. Chiang et al.]
{Chia-Ying Chiang$^{1}\thanks{E-mail: cychiang@ast.cam.ac.uk}$, R. C. Reis$^{1,2}$, A. C. Fabian$^{1}$, D. Grupe$^{3}$ and S. Tsuruta$^{4}$\\
$^1$Institute of Astronomy, University of Cambridge, Madingley Road, Cambridge
CB3 0HA\\
$^2$Department of Astronomy, University of Michigan, Ann Arbor, MI
48109, USA\\
$^3$Department of Astronomy and Astrophysics, Pennsylvania State University, 525 Davey Lab, University Park, PA 16802, USA\\
$^4$Department of Physics, Montana State University, Bozeman, MT 59717, USA\\
}

\date{Accepted 2012 June 15. Received 2012 June 14; in original form 2011 October 25}
\pagerange{\pageref{firstpage}--\pageref{lastpage}} \pubyear{2012}

\begin{document}

\topmargin = -0.5cm

\maketitle

\label{firstpage}

\begin{abstract}
We analysed new simultaneous \emph{Suzaku} and \emph{Swift} data
of the Broad Line Seyfert 1 (BLS1) galaxy CBS 126. A clear Fe
emission line and a strong soft excess are present in the source
spectra. We fit the spectra with a relativistic reflection model and
find the model tends to fit the data with a high iron abundance
possibly due to the large soft excess present. By checking the
difference and the RMS spectra, we find there is likely an
absorption edge at $\sim$ 0.89 keV, which might be caused by oxygen
or neon. We also produced an analysis of the time-resolved spectra in order
to examine the existence of the edge. Although high
iron abundance is required more in the time-resolved spectra, a model
of solar iron abundance together with an absorption edge gives a more
physical explanation. Variation of the ionisation parameter is an
alternative, plausible explanation for the excess seen in the
difference spectra. It is difficult to know if there are warm
absorbers in this source from the current data. To further
investigate the presence of possible warm absorbers, higher
signal-to-noise low-energy data are needed. The work presented here
tentatively suggests that the spectra of a BLS1 can be explained by
a relativistic reflection model similar to that often seen in their
narrow-line cousins.

\end{abstract}


\section{Introduction}

Broad-line Seyfert 1 galaxies (BLS1s) are a subclass of Active
Galactic Nuclei (AGN) with broad permitted optical emission lines
(H$\beta$ FWHM $\geq$ 3000 km s$^{-1}$). Unlike narrow-line Seyfert
1 galaxies (NLS1s; H$\beta$ FWHM $<$ 2000 km s$^{-1}$), the
broad-line region (BLR) of a BLS1 is of high velocity and close to
the central black hole. Studies of \citet{Boller96} concluded that
high-velocity and steep soft X-ray continuum slopes are not found in
nature in AGN population. The steep, ultrasoft X-ray spectra in
NLS1s are generally believed to be because of a relatively low-mass
black hole with a very high accretion rate \citep{Pounds95} or
distant BLR clouds \citep{Puch92}.

%a low Galactic absorption column ($N_{\mbox{\scriptsize H}} = 1.38 \times 10^{20}$ cm$^{-2}$, Dickey \& Lockman).

\begin{figure}
\includegraphics[scale=0.4]{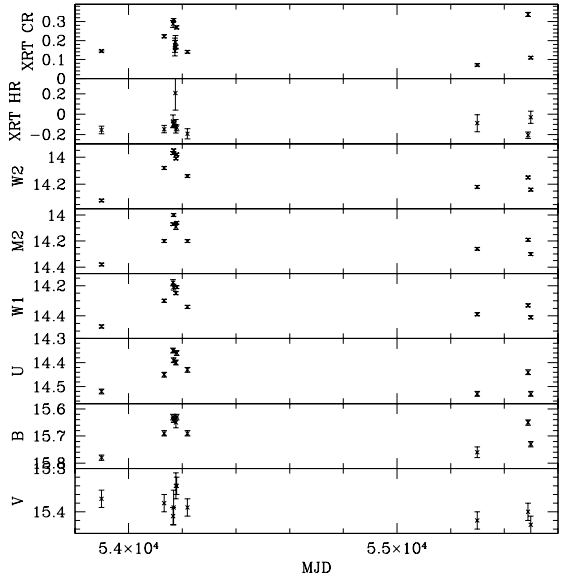}
\caption{The figure shows the light curves of \emph{Swift}
instruments. The top two panels show the XRT count rate and hardness
ratio, and the other panels show data extracted from different bands
of UVOT data. We define the hardness ratio as HR=(H-S)/(H+S) where S
and H are the  background-corrected number of counts in the 0.3-1.0
and 1.0-10.0 keV energy bands, respectively.} \label{lc_swift}
\end{figure}

CBS 126 (= Ton 1187, $\alpha_{2000}$ = 10h 13m 03.2s, $\delta_{2000}
= +35^{\circ} 51^{'} 24^{''}$, z = 0.079)
 is a broad line Seyfert 1 galaxy with H$\beta$ = 2980 $\pm$ 200 km s$^{-1}$ \citep{Grupe04}.
% I left out the galactic absorption here because it is not an intrisic property of the source. This information
% needs to to into the spectral analysis section.
It was detected as a bright X-ray AGN during the \textit{ROSAT}
All-Sky Survey (RASS, \citealt{Voges99}) in 1990 November
\citep{Grupe98} and was observed again by \textit{ROSAT} with the
High Resolution Imager in 1995 October \citep{Grupe01}. Our
attention was drawn to CBS 126 because it shows a relatively high
degree of optical continuum polarisation. \citet{Grupe98b} measured
a degree of polarisation  1.26$\pm0.13$ per cent, which is quite unusual
for a soft X-ray selected AGN. The presence of optical polarisation
suggests that part of the direct view is absorbed/reddened.
\citet{Brandt97} suggested and found that soft X-ray selected,
highly polarised AGN show warm, ionised absorber in X-rays. CBS 126
is therefore a good candidate for finding ionised X-ray absorber
features. It was again observed in X-rays as part of a \emph{Swift}
fill-in project to study the spectral energy distributions of AGN
\citep{Grupe10} in 2006 June. The source shows similar properties to
NLS1s such as strong spectral variability (see Fig. \ref{lc_swift})
and a large soft excess below $\sim$ 1 keV, which are typical for
NLS1s but unusual for BLS1s. CBS 126 harbours a black hole with an
estimated mass in excess of $\sim 7.6 \times 10^{7}$ M$_{\odot}$
\citep[ SDSS DR7 catalogue]{Shen11}. This, together with the
high-velocity BLR clouds and the enormous soft excess make CBS 126 an
interesting source.
% leave out the last sentence about the BLR and the soft excess. Note that the BLR as far as we know today does not consists
% of clouds. Also the soft excess is not enormous. There are even more extreme sources with that respect (see my 2010 paper)

A soft excess below $\sim$ 1 keV is
commonly seen in X-ray spectra of AGNs. It can be fitted phenomenologically by a
blackbody with a temperature of $\sim$ 0.1 - 0.2 keV, but the
temperature of the component is fairly constant between sources, and is probably not
related to the luminosity or black hole mass of the source
\citep{Gierlinski04}. Instead, it is likely to have an atomic
origin. Since the disc model predicts that the disc temperature
should be correlated with the black hole mass, the soft excess is
less likely to be due to the thermal emission from the accretion
disc, except for objects with low black hole masses and high Eddington
ratios such as NLS1s. \citet{Gierlinski04,Gierlinski06} suggested that soft excess
can be caused by relativistically smeared and partially ionised
absorption, but in this scenario an extreme outflow velocity is
needed to produce a smooth spectrum \citep{Schurch07}.
\citet{Done12} claimed that if the inner disc emission 
emerges as a Comptonisation component rather than a blackbody
spectrum, the Wien tail of this component could produce the 
soft excess. \citet{Jin12} showed that the model could 
explain the spectra of a sample of unobscured type 1 AGNs. This
scenario requires an electron population with a relatively low temperature and
high optical depth, and the Compton upscattering component
to arise in the transition region between the disc and the corona. It again
requires an unexplained, fairly constant temperature between sources.

An alternative physical interpretation to explain the broad-band X-ray 
spectrum is the relativistic reflection model.
\citet{Crummy06} investigated a large sample of type 1 AGNs and
found the relativistically blurred reflection can reproduce
the soft excess. The soft excess here is a series of blurred 
emission lines caused by disc reflection and other emission
\citep{Ross05}. The constancy of the shapes is then due to atomic
physics.

One of the major goals in observing CBS 126 with \emph{Suzaku} and
\emph{Swift} simultaneously, was to clarify the nature of its
unusual, variable, optically polarised spectrum as found from
earlier \emph{Swift} observations. Fig. \ref{lc_swift} exhibits
correlated variability in the optical/UV and X-ray bands. Most
importantly, the earlier Swift data already showed that the
variability in the X-ray band itself takes place mostly at higher
energies, while at the lower X-ray energies there was no significant
variability. Furthermore, earlier \emph{Swift} data exhibited
 steep soft X-ray spectra during dimmer
states while flatter spectra during higher states.
Standard accretion disc-corona models cannot explain this kind of
behaviour. For instance, those models would expect the soft X-rays
to vary in a similar manner to the hard X-rays, i.e. the variety
would be broadband in X-rays, which is not the case with CBS 126. On
the other hand, the spectral and temporal behaviour of this source
can be {\it well explained} with ionised relativistic reflection
models \citep{Miniutti04}. In this picture the relativistic
reflection component, including the soft X-ray excess, dominates
during dimmer states, and thus give rise to the observed steep soft
X-ray spectra. In contrast, during brighter states, the reflection
component is diluted and the flatter powerlaw dominates. The focus
of this paper is to test whether our new simultaneous
\emph{Suzaku}/\emph{Swift} data further support the relativistic
reflection model for CBS 126. Other possibilities will not be
probed in depth.

\section{Data Reduction} \label{data_reduction}
\subsection{\emph{Suzaku}} \label{data_Suzaku}
\begin{figure}
\leavevmode \epsfxsize=8.5cm \epsfbox{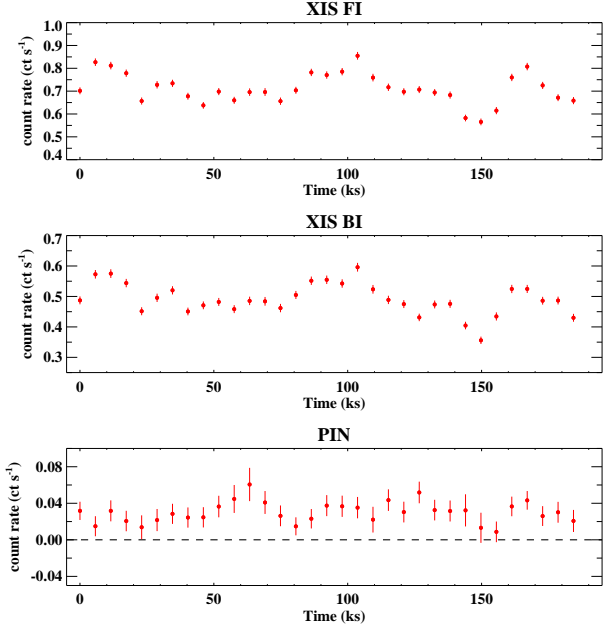} \caption{The figure
shows the background-corrected light curves of FI and BI XIS
detectors, and PIN data. These light curves have been binned to
orbital bins (5760 s) for clarity.} \label{lc_all}
\end{figure}

CBS 126 was observed by \emph{Suzaku} on 2010 October 18 for a total
of $\sim$100 ks of good exposure time (see Table \ref{suzaku_log}).
The X-ray Imaging Spectrometer (XIS) was operated in the normal
mode. All the detectors (XIS0, XIS1 and XIS3) were operated in both
$3\times3$ and $5\times5$ editing modes. We reduced the data with
the {\sevensize HEASOFT V6.10} software package following the Suzaku
Data Reduction Guide. We used a circle with a 180 arcsec diameter to
extract the source spectrum. The background spectrum was extracted
using the same size of circular region from a position free from the
source, the calibration source and any other clear cosmic source.
Response files were produced by the script \texttt{XISRESP}, which
calls \texttt{XISRMFGEN} and \texttt{XISSIMARFGEN} automatically.
Spectra and response files of the two front-illuminated (FI) CCD XIS
detectors (XIS0 and XIS3) were combined using the script
\texttt{addascaspec} in {\sevensize FTOOL}, resulting in a total 2 -
10 keV flux of which is 4.12 $\times 10^{-12}$ ergs cm$^{-2}$
s$^{-1}$. The XIS 0.5-10.0 keV light curve of CBS 126 is shown in
the upper panel in Fig. \ref{lc_all}. It is clear that the XIS light
curve is variable. We also produced the difference spectrum by
subtracting the low-flux state spectrum from the high-flux state
spectrum. The XIS light curve was divided into two flux intervals,
both having a similar number of counts. The slice above the average
count rate (1.25 ct/s for XIS FI and 2.00 ct/s for XIS BI) forms the
high-flux state spectrum, while the slice below the average
represents the low flux spectrum. The total 0.5 - 2 keV fluxes are
4.57 $\times 10^{-12}$ ergs cm$^{-2}$ s$^{-1}$ and 4.49 $\times
10^{-12}$ ergs cm$^{-2}$ s$^{-1}$ for XIS FI and back-illuminated
(BI, that is, XIS1) spectra, respectively.

\begin{table}
  \centering
  \caption{Summary of the \emph{Suzaku} observations of CBS 126.
  The 0.5-10 keV net source count rates for different detectors are listed below.
  }
  \label{suzaku_log}
  \begin{tabular}{cccrc}
  \hline
  \hline

Detector &  window & MJD  & Exposure & Count Rate \\
\hline
XIS0 & 3 $\times$ 3 & 55487.358  &  72505  &  $0.367\pm0.002$   \\
        & 5 $\times$ 5 & 55487.831  &  29061  &  $0.339\pm0.003$    \\
XIS1 & 3 $\times$ 3 & 55487.358  &  72481  &  $0.811\pm0.003$    \\
        & 5 $\times$ 5 & 55487.831  &  29061  &  $0.743\pm0.005$    \\
XIS3 & 3 $\times$ 3 & 55487.358  &  72505  &  $0.404\pm0.002$    \\
        & 5 $\times$ 5 & 55487.831  &  29061  &  $0.378\pm0.004$     \\
\hline
\hline
\end{tabular}
\end{table}

\begin{figure}
\leavevmode \epsfxsize=8.5cm \epsfbox{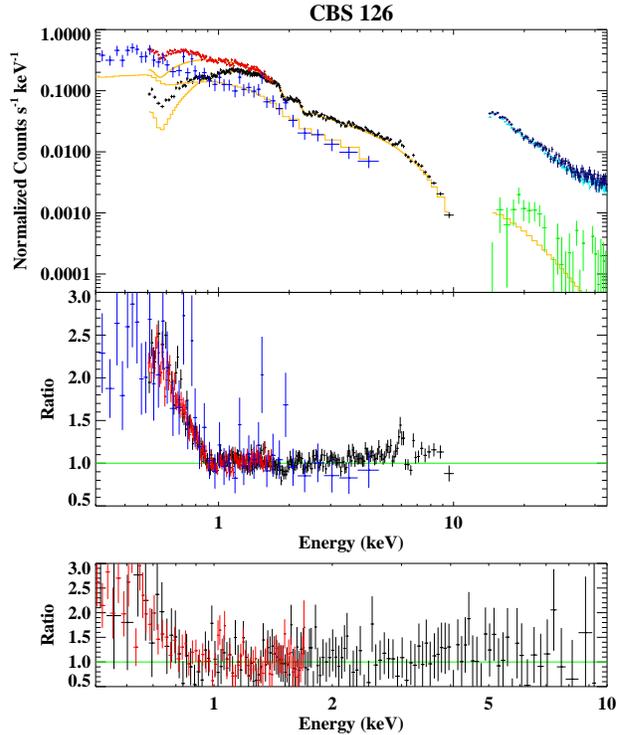} \caption{
The top panel shows the overall spectra fitted with
a powerlaw across the 0.9 to 4 keV and 7.5 to 10 keV energy range.
We show the FI XIS and BI XIS spectra in black and red
respectively. The PIN spectrum is shown in green and the
\emph{Swift} XRT spectrum is shown in blue. The navy, blue and cyan
data are explained in detail in section \ref{data_Suzaku}. A strong
soft excess at low energies and iron line structure can be clearly
seen. The lower panel shows the data/model ratio of the
difference spectra (see definition in section \ref{data_Suzaku}) with the same powerlaw model applied. The BI XIS
difference spectrum above 1.7 keV and the FI XIS
above 10 keV are ignored due to poor signal-to-noise.} \label{pow}
\end{figure}

The Hard X-ray Detector (HXD) was operated in XIS-nominal pointing
mode. The background spectrum of HXD/PIN consists of a non-X-ray
background and a cosmic X-ray background. We obtained the non-X-ray
background event file from a database of background observations
made by the PIN diode. The background model D (the tuned model) was
used for non-X-ray background extraction. The cosmic X-ray
background was estimated by model simulation using the PIN response
for flat emission distribution, and the count rate in this case is
$0.187 \pm 0.001$ counts s$^{-1}$ in the effective PIN energy band
14.0-45.0 keV (all the count rates of PIN will be given over this
band if not specified). The source is faint ($0.328 \pm 0.002$
counts s$^{-1}$), appearing at only $\sim$ 4 percent above the
background level ($0.316 \pm 0.001$ counts s$^{-1}$). Since the
background dominates the PIN spectrum, we produced the
earth-occulted background for a further check. Ideally the
earth-occulted background should be identical to the non-X-ray
background. The counts of the earth-occulted background ($0.358 \pm
0.004$ counts s$^{-1}$) are however slightly higher ($\sim$ 15
percent) than that of the non-X-ray background ($0.298 \pm 0.001$
counts s$^{-1}$), which means the PIN data were mildly
over-estimated. The navy blue points in Fig. \ref{pow} is the total
background including the earth-occulted background and the cosmic
background, and the cyan points slightly below stand for the PIN
spectrum before background correction. It can be seen that if we
replace the non-X-ray background with the earth-occulted background,
we effectively have a non-detection in the PIN band. The green
points in Fig. \ref{pow} are the background-corrected PIN data when
using the non-X-ray background. Since the data show no solid
detection in PIN, we do not include the PIN spectrum in all our
spectral fittings. The light curve plotted in the lower panel of
Fig. \ref{lc_all} also shows the source is faint in the PIN energy
band, thus we did not generate a difference spectrum for PIN either.

\begin{table*}
  \centering
  \caption{Summary of the \emph{Swift} observations of CBS 126.
  Start and end times $T_{\rm start}$ and $T_{\rm end}$ of the observations are
  given in UT and all exposure times are given in units of s. XRT Count rates and hardness ratios (HR) of CBS 126 are also displayed.
  The UVOT magnitudes were corrected for Galactic reddening \citep[$E_{\rm B-V}$ = 0.011;][]{Schlegel98}.
  }
  \label{obs_log}
  \begin{tabular}{cccrcccccccc}
  \hline
  \hline

Segment &  $T_{\rm start}$  & MJD & $T_{\rm XRT}$  & Count Rate & HR & V & B & UV W1 & UV W2 \\
\hline
001 & 2006-06-13 00:18 & 53899.010 & 4725 & 0.144$\pm0.006$ & $-$0.16$\pm0.04$ & $15.37\pm0.02$ &  $15.78\pm0.01$  &  $14.47\pm0.01$ &  $14.32\pm0.01$ \\
002 & 2007-02-01 11:29 & 54132.476 & 4014 & 0.222$\pm0.008$ & $-$0.15$\pm0.04$ & $15.38\pm0.02$ &  $15.69\pm0.01$ &  $14.30\pm0.01$   &  $14.08\pm0.01$ \\
003 & 2007-03-06 12:01 & 54165.499 &  913 &  0.288$\pm0.018$ & $-$0.07$\pm0.06$ & $15.41\pm0.02$ & $15.63\pm0.01$ &  $14.18\pm0.02$ &  $13.97\pm0.01$  \\
004 & 2007-03-08 13:18 & 54167.552 & 4977 &  0.307$\pm0.009$ & $-$0.09$\pm0.03$ &  $15.39\pm0.04$ &  $15.64\pm0.01$    & $14.21\pm0.02$ &  $13.95\pm0.01$\\
007 & 2007-03-13 07:58 & 54172.331 &   99 &   0.160$\pm0.040$ & ---         &  ---  &   ---  &   ---  &   ---    \\
008 & 2007-03-14 01:39 & 54172.999 &  125 &  0.175$\pm0.038$ & ---         &  ---  &   ---   &   ---  &   ---   \\
009 & 2007-03-15 14:36 & 54174.610 &  202 &   0.193$\pm0.033$ & $+$0.21$\pm0.16$ &  ---  &   ---  &   ---  &   ---    \\
010 & 2007-03-17 08:05 & 54176.335 & 1782 &   0.166$\pm0.011$ & $-$0.12$\pm0.07$ & $15.34\pm0.03$ &  $15.65\pm0.02$  &  $14.25\pm0.01$ &  $14.01\pm0.01$\\
011 & 2007-03-20 08:27 & 54179.351 & 4266 &   0.269$\pm0.008$ & $-$0.13$\pm0.03$ & $15.34\pm0.02$ &  $15.63\pm0.01$  &  $14.21\pm0.01$ &  $13.98\pm0.01$\\
015 & 2007-04-29 08:55 & 54219.369 & 3069 &  0.141$\pm0.007$ & $-$0.19$\pm0.05$ & $15.39\pm0.02$ &  $15.69\pm0.01$  &  $14.34\pm0.01$  &  $14.14\pm0.01$\\
016 & 2010-04-13 02:15 & 55299.092 & 2608 &   0.072$\pm0.006$ & $-$0.09$\pm0.08$ & $15.42\pm0.02$ & $15.76\pm0.02$ &  $14.39\pm0.01$  &  $14.22\pm0.01$\\
017 & 2010-10-19 08:30 & 55488.353 & 3160 &   0.337$\pm0.011$ & $-$0.21$\pm0.03$ & $15.40\pm0.02$ & $15.65\pm0.01$ &  $14.33\pm0.01$  &  $14.15\pm0.01$\\
018 & 2010-10-29 01:14 & 55498.162 & 2986 &   0.111$\pm0.007$ & $-$0.03$\pm0.06$ & $15.43\pm0.02$ &  $15.73\pm0.01$  &  $14.41\pm0.01$ &  $14.24\pm0.01$\\
\hline
\hline
\end{tabular}
\end{table*}

\subsection{\emph{Swift}}
The \emph{Swift} Gamma-Ray Burst (GRB) explorer mission
\citep{Gehrels04} observed  CBS 126 13 times. Note that some of
these observations were very short due to \emph{Swift} triggers on
Gamma Ray Bursts. A summary of all \emph{Swift} observations between
2006 June 13  and 2010 October 29 is given in Table \ref{obs_log}.
We use \emph{Swift} data taken simultaneously with the above
\emph{Suzaku} observation. The \emph{Swift} X-Ray Telescope (XRT;
\citealt{Burrows05}) was operating in photon counting mode (PC mode;
\citealt{Hill04}) and the data were reduced by the task {\it
xrtpipeline} version 0.10.4, which is included in the {\sevensize
HEASOFT} package 6.1. Source photons were selected in a circular
region  with a radius of 70$^{''}$ and background region of a close
by source-free region with $r=200^{''}$. The background region is
much larger than the source region in order to get enough counts to
model the background spectrum. Photons were selected with grades
0--12. The photons were extracted with {\sevensize XSELECT} version
2.4. The spectral data were re-binned by using {\it grppha} version
3.0.0 having 20 photons per bin. The spectra were analysed with
{\sevensize XSPEC} version 12.3.1 \citep{Arnaud96} between 0.3 and 5
keV. The ancillary response function files (arfs) were created by
{\it xrtmkarf} and corrected for vignetting and bad columns/pixels
using the exposure maps. The standard response matrices {\it
swxpc0to12s0\_20010101v011.rmf} and {\it
swxpc0to12s6\_20010101v011.rmf} were used for the observations
before and after the XRT substrate voltage change in 2007 August,
respectively \citep{Godet09}. The difference spectrum was not
generated because the exposure time ($\sim 6$ ks) is too short.

\begin{figure}
\leavevmode \epsfxsize=8.5cm \epsfbox{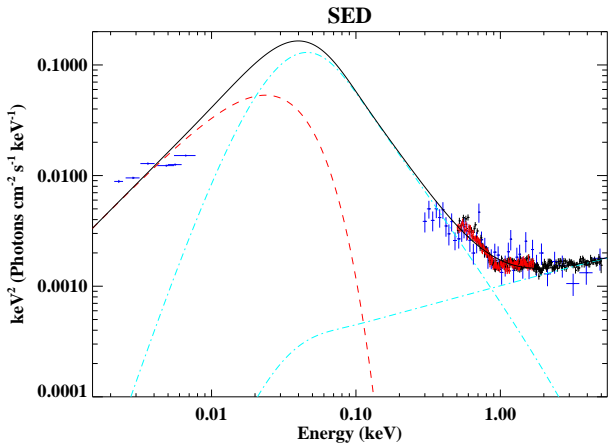}
\caption{The figure shows the SED plot from the 17 (blue points)
segment of \emph{Swift} data which is simultaneous with our
\emph{Suzaku} observation. \emph{Suzaku} XIS FI (black) and BI (red)
data have been included for reference. The optical-X-ray connecting
index $\alpha_{\rm{ox}}$ is $\sim$ 1.42.} \label{sed}
\end{figure}

One of the advantages of \emph{Swift} is that data are taken
simultaneously in X-rays and in the Optical/UV. Therefore, data were
also taken with the UV/Optical Telescope (UVOT; \citet{Roming05})
using 6 photometry filters. Before analysing the data, the exposures
of each segment were co-added by the UVOT task {\it uvotimsum}.
Source counts were selected with the standard 5$^{''}$ radius in all
filters \citep{Poole08} and background counts in a source-free
region with a radius r=20$^{''}$. The data were analysed with the
UVOT software tool {\it uvotsource} assuming a GRB like power law
continuum spectrum. The magnitudes as listed in Table \ref{obs_log}
were all corrected for Galactic reddening $E_{\rm B-V}=0.011$ given
by \citet{Schlegel98}. The correction factor in each filter was
calculated with equation (2) in \citet{Roming09} who used the
standard reddening correction curves by \citet{Cardelli89}.

\begin{figure}
\leavevmode \epsfxsize=8.5cm \epsfbox{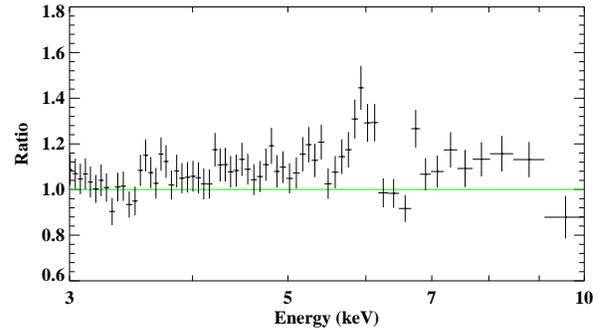} \caption{The
figure shows ratio plot of the iron line profile of FI
XIS spectrum. The equivalent width (EW) of the line is $114^{+35}_{-36}$
eV. 
} \label{zoomiron}
\end{figure}

\section{Data Analysis}

We fit a powerlaw modified with Galactic absorption across the 0.9-4
keV and 7.5-10 keV energy ranges, and the results are shown in Fig.
\ref{pow}. A large soft excess in both FI and BI XIS spectra and the
\emph{Swift} XRT spectrum, and an iron emission line in the FI XIS
spectrum can be clearly seen. We also fit the modified powerlaw
model to the \emph{Suzaku} difference spectra (see the lower panel
of Fig. \ref{pow}). The difference spectra are quite noisy due to
lack of data. A simple powerlaw fits the difference spectra above 1
keV, and there are no features around the iron line band as expected
if the features are due to constant, additive component such as an
emission line. Despite poor signal-to-noise in the difference
spectrum, soft excesses are still present, similar to that seen in
the real spectra. This hints at the possibility of either a further
distinct varying component acting preferentially in the low-energy
band or the presence of a multiplicative component, as the latter is
not removed in the difference spectra even if it does not change
with time.

In the frame work of an extra component acting preferentially at low
energies, the large soft excess could be due to a possible
Comptonised Wien tail from the disc component. We investigate the
possibility of this scenario by fitting the spectral energy
distribution (SED) of the UV and soft X-ray data of CBS 126 with a
Comptonisation model. The UV data have been fitted with a
low-temperature blackbody disc component ({\sevensize DISKBB},
\citealt{diskbb}). Two Comptonisation components ({\sevensize
COMPTT}, \citealt{comptt}) have been included to explain the soft
excess and the powerlaw-shape spectrum at high energies. As shown in
Fig. \ref{sed}, the Comptonised Wien tail can reproduce the shape of
the soft excess. If the component which produces the soft excess varies
with the hard powerlaw-shape component, the model could explain
both the real and difference spectra.
However, the possible iron emission feature present in the 
\emph{Suzaku} FI XIS spectrum (see Fig. \ref{zoomiron}) with a line width of $0.12^{+0.07}_{-0.05}$ keV at $\sim$ 6.41 keV (rest energy), 
though not significantly broadened, together with the large soft excess which could be caused by blurred reflection, act to validate
the relativistic reflection scenario. As stated in the introduction, 
the purpose of the work is to test if the relativistic reflection model
can explain the overall shape of the X-ray continuum of the new \emph{Suzaku} observation. In what follows, 
we explore this possibility further, though it may not be the only
explanation of the data as the Compton hump has not been detected by
\emph{Suzaku} PIN.

%for it to account for the excess ALSO
%observed in the difference spectrum, the overall disc flux would 
%have to be varying at very short time scales, which is not a likely
%physical possibility for such a supermassive black hole. A further,
%in our opinion more likely, possibility that can explain the soft
%excess in BOTH the real as well as difference spectra, is that it is
%caused by relativistically blurred reflection. 

\begin{table}
 \caption{The table below lists the fitting parameters and $\chi^{2}$ obtained
by models fitted over different energies. $\Gamma$ is the photon
index of the powerlaw component, and $N_{\rm pow}$ stands for the
normalisation of the powerlow component. $\tau$ is the depth of the
absorption edge. ``Index" is for the emissivity profile index in the
convolution model {\sevensize KDBLUR}. $N_{\rm iref}$ and
$N_{\rm nref}$ represent the normalisations of the ionised and
neutral reflection components, respectively. In this table all the
errors are at the 90\% confidence level.
 }
\label{fitting}
\begin{tabular}{@{}lcccc}
\hline\hline
 & \multicolumn{2}{c}{Time-averaged} & \multicolumn{2}{c}{Time-resolved}\\
 & Model A & Model B & Model A & Model B\\
\hline
$\Gamma$ & $2.01\pm0.01$ & $2.16^{+0.01}_{-0.02}$ & $2.00^{+0.02}_{-0.03}$ & $2.11^{+0.03}_{-0.02}$\\
$N_{\rm pow}$ ($10^{-3}$) & $1.5\pm0.1$ & $1.4\pm0.1$ & \multicolumn{2}{c}{See Table 4}\\
\hline
$E_{\mbox{\scriptsize edge}}$ & - & $0.89\pm0.01$ & - & $0.88^{+0.02}_{-0.01}$\\
$\tau$ & - & $0.35^{+0.02}_{-0.03}$ & - & $0.39^{+0.05}_{-0.04}$\\
\hline
index & $7.2^{+0.8}_{-1.3}$ & $> 5.4$   & $4.2^{+0.3}_{-0.1}$ & $5.8\pm2.1$\\
$R_{\mbox{\scriptsize in}}$(R$_{\mbox{\scriptsize g}}$) & $3.0^{+0.6}_{-0.4}$ & $1.7^{+0.6}_{-0.1}$ & $<$ 2.2 & $1.7^{+0.5}_{-0.2}$\\
$i$ & $27^{+20}_{-6}$ $^{\circ}$ & $51^{+3}_{-14}$ ${^\circ}$ & $<$ 19 $^{\circ}$ & $51^{+11}_{-16}$ $^{\circ}$\\
\hline
$A_{\mbox{\scriptsize Fe}}$ & $>$ 9.0 & (1.0) & $>$ 9.5 & (1.0)\\
$\xi$ & $200^{+12}_{-23}$ & $119^{+33}_{-28}$ & \multicolumn{2}{c}{See Table 4}\\
$N_{\rm iref}$ ($10^{-7}$) & $2.5^{+0.6}_{-1.2}$ & $6.2^{+3.4}_{-0.9}$ & \multicolumn{2}{c}{See Table 4}\\
$N_{\rm nref}$ ($10^{-6}$)& $8.9^{+2.6}_{-2.5}$ & $10.7^{+3.8}_{-4.1}$ & $8.7\pm2.3$ & $11.0^{+5.1}_{-4.8}$\\
\hline
$\chi^{2}/d.o.f.$ & 1642/1476 & 1632/1475 & 1148/1067 & 1181/1066\\
% & (1.12) & (1.13) & (1.08) & (1.11)\\
\hline\hline
\end{tabular}
\end{table}

\subsection{Time-averaged Spectra}

We start by fitting the data with the simplest model that can explain the broad
iron line in a physically consistent manner as well as the soft excess. The model consists of a
reflection continuum model {\sevensize REFLIONX} \citep{Ross05}
convolved by the {\sevensize KDBLUR} model. We also model the
neutral reflection with {\sevensize REFLIONX} by setting the
ionisation parameter $\xi = 1.0$. The Galactic absorption is
modelled by {\sevensize TBnew} using Wilms \citep{Wilms00}
abundance. The model (hereafter Model A) and the fitting parameters
are summarised in Table \ref{fitting}.

\begin{figure}
\leavevmode \epsfxsize=8.5cm \epsfbox{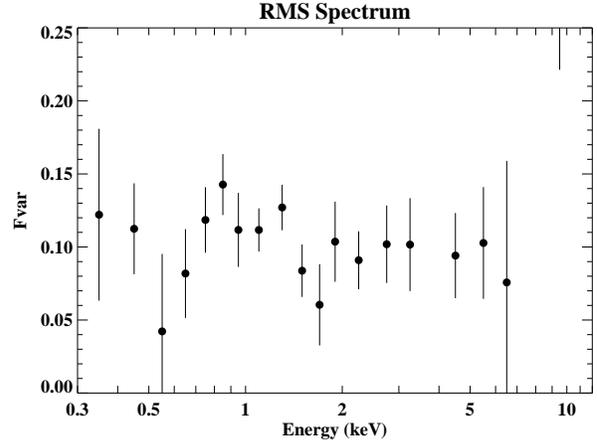}
\caption{The figure shows the RMS spectrum of CBS 126. Each data
point is calculated using a light curve with a time bin of 1ks. The
last data point around 10 keV shows a much higher fraction of
variability than the others possibly due to severe noise at the highest energy
band.} \label{rms}
\end{figure}

\begin{figure}
\includegraphics[scale=0.38]{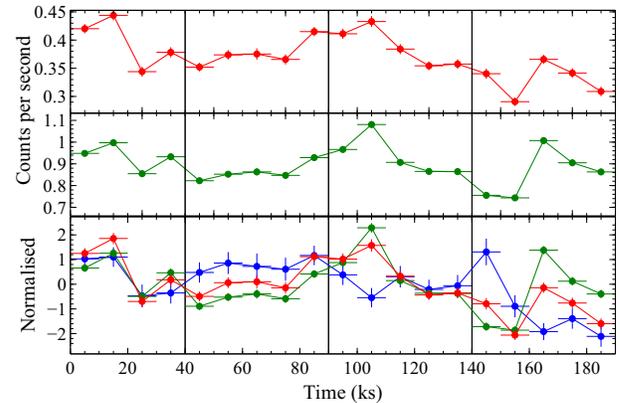}
\caption{
XIS light curve in the hard (1.0-10.0 keV; red ) and soft (0.3-1.0 keV; green) energy band and colour evolution
(soft/hard; blue). In the lower panel these were normalized by dividing the difference
between the count rate in each 10 ks bin and the mean count rate by
the standard deviation. See section \ref{TRS} for detail.} \label{lc_time}
\end{figure}

It is found that the iron abundance $A_{\mbox{\scriptsize Fe}}$
tends to be higher than 9 times of the solar value, which is not
expected. If we set the iron abundance to be the same as the solar
value, the $\chi^{2}/d.o.f.$ of the fitting increases to 2014/1477
and the model failed to reproduce the soft excesses below 0.9 keV. A
blackbody component possibly arising from a low-temperature
accretion disc was then considered, even though we do not expect the
disc emission to change within such a short period of time and thus
should not give rise to the soft excess seen in the difference
spectrum. To examine whether the disc emission is present in the
X-ray band and contributes variability, we generated the root mean
square (RMS) spectrum using the technique developed by
\citet{Edelson02}. The RMS spectrum of CBS126 (see Fig. \ref{rms})
is flat, and the variability of each energy band lies on a similar
level. This implies that not a particular energy band contributes
significantly more variability than the other energy bands.
Furthermore, when adding a {\sevensize DISKBB} component to the
spectra, an effective temperature $T_{\rm eff}$ of $\sim$ 50 eV is
required to explain the soft excess, which is way too high for a $M
\sim 10^8 M_{\odot}$ black hole (typical value $T_{\rm eff} \sim$ 10
eV for black holes of this mass). Therefore, the possibility of a
blackbody component can be ruled out and either the source has a
highly supersolar iron abundance or there is a further component
contributing to the soft excess which must also be present in the
difference spectrum.

\begin{table}
 \caption{The table below lists the fitting parameters of different periods of the observation obtained by the time-resolved
 analysis. Definitions of these parameters are identical with those
 shown in Table 3.
 }
\label{time_resolved}
\begin{tabular}{@{}lcccc}
\hline\hline
Time-resolved & P1 & P2 & P3 & P4\\
\hline
 & \multicolumn{4}{c}{Model A}\\
$N_{\rm pow}$ ($10^{-3}$) & $1.5\pm0.1$ & $1.4\pm0.1$ & $1.4\pm0.1$ & $1.3\pm0.1$\\
$\xi$ & $200^{+11}_{-42}$ & $193^{+15}_{-49}$ & $109^{+14}_{-6}$ & $104^{+8}_{-11}$\\
$N_{\rm iref}$ ($10^{-7}$) & $3.1^{+1.2}_{-0.6}$ & $3.3^{+1.0}_{-0.6}$ & $12.1^{+3.3}_{-5.0}$ & $12.1^{+1.9}_{-7.5}$\\
\hline
 & \multicolumn{4}{c}{Model B}\\
$N_{\rm pow}$ ($10^{-3}$) & $1.5\pm0.1$ & $1.3\pm0.1$ & $1.5\pm0.1$ & $1.4\pm0.1$\\
$\xi$ & $208^{+28}_{-61}$ & $208^{+24}_{-98}$ & $103^{+34}_{+32}$ & $90^{+31}_{-29}$\\
$N_{\rm iref}$ ($10^{-7}$) & $2.3^{+0.9}_{-0.6}$ & $2.4^{+1.7}_{-0.6}$ & $7.6^{+5.4}_{-3.6}$ & $6.9^{+2.7}_{-1.2}$\\
\hline\hline
\end{tabular}
\end{table}

In order to test the possibility that the soft excess in the
difference spectrum is due to a combination of reflection at solar
abundances and a further multiplicative component, we added an
absorption edge using {\sevensize ZEDGE} to Model A and froze the
iron abundance at solar. This model is hereafter referred to as
Model B (Table \ref{fitting}). This resulted in a fit with similar
quality ($\chi^{2}/d.o.f.$ = 1632/1475) to that of Model A when the
iron abundance was $>$ 9. The energy of the absorption edge is
$\sim$ 0.89 keV, which is likely associated with oxygen or neon.

It is clear from Fig. \ref{lc_all} that the \emph{Suzaku} XIS light
curves are variable. In the following we extend our spectral
analysis to time-resolved spectra.

\subsection{Time-resolved Spectra} \label{TRS}
\subsubsection{Spectral Fitting}
A 0.3-1.0 keV and a 1.0-10.0 keV \emph{Suzaku} XIS light curves with
a 10 ks time bin were created and are shown in Fig. \ref{lc_time}.
The top two panels of Fig. \ref{lc_time} shows the ``soft" (0.3-1.0
keV) and the ``hard" (1.0-10.0 keV) light curves, respectively. The
blue line in the lower panel shows the colour (the soft/hard ratio)
evolution of the observation. When we look at the top two panels of
Fig. \ref{lc_time}, the count rates of the soft and the hard light
curves appear to vary with each other. However, looking at the
normalised colour evolution (soft/hard), it appears that there could
in fact be at least four distinct periods which are highlighted by the
vertical lines in Fig. \ref{lc_time}. As such, we proceed by dividing
the observation into these different segments and created
a new set of time-resolved spectra. The observation was divided into
the segments as follows:
0-40 ks (period 1), 40-90 ks (period 2), 90-140 ks (period 3) and
140-190 ks (period 4). The resulting \emph{Suzaku} FI XIS spectra
are noisy in the low-energy band (of particular interest to us since
we are looking for the soft excess), therefore only the \emph{Suzaku}
low-energy BI XIS spectra were used. The \emph{Swift} XRT spectrum has an
exposure time much less than 40 ks and is not suitable for
time-resolving analysis envisioned here.

\begin{figure}
\leavevmode \epsfxsize=8.5cm \epsfbox{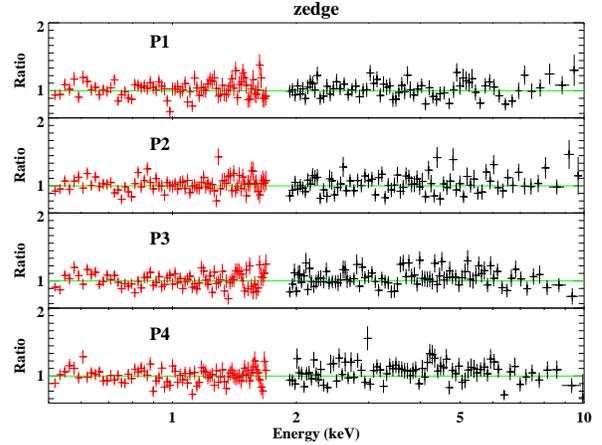} \caption{The
figure shows the fitting results of spectra in different periods
using model B. The black and red points represent FI and BI XIS
data, respectively. The FI XIS spectra below 1.9 keV were ignored
because the spectra turned to be noisy after breaking down into
segments.} \label{ze_ra}
\end{figure}

We fit each time-resolved spectrum with both models A and B. The
iron abundance, inclination, neutral hydrogen density, emissivity
index, edge energy and inner radius are treated as global parameters
and are tied between the periods (i.e., they are assumed to be
constant at these time scales). The normalisation of the powerlaw
and ionised reflection component, as well as the ionisation
parameter of the inner disc, on the other hand, are free to vary.
The normalisation of the neutral reflection component is also free
to vary, but tied between the periods. Table \ref{fitting} also
summarises the results for the global parameters in the
time-resolved data  and Table \ref{time_resolved} shows the
evolution of the remaining parameters. Fig. \ref{ze_ra} displays the
data/model ratio of  for Model B, with the  different periods shown
in separate panels for clarity. It can be seen that that this model
fits the spectra of each periods well, with no evidence for a soft
excess. The energy of the absorption edge is found to be $\sim$ 0.88
keV, which is consistent with that obtained by the time-averaged
spectra.

\begin{figure}
\leavevmode \epsfxsize=8.5cm \epsfbox{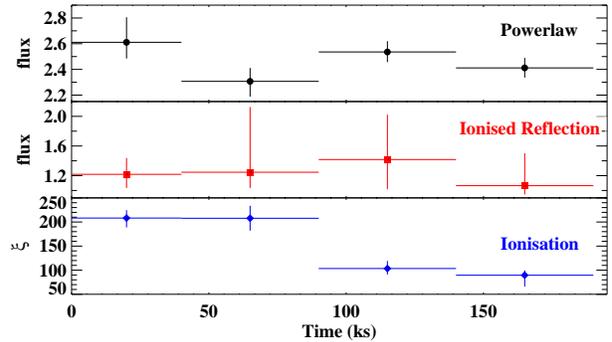}
\caption{The figure shows, from top to bottom, the 0.01-1000 keV flux (in unit of
$10^{-11}$ erg cm$^{-2}$ s$^{-1}$) evolution of the powerlaw
component, the ionised reflection component, and the evolution of
the ionisation state. In this figure the error bars are of
1$\sigma$.} \label{flux_evo}
\end{figure}

\begin{figure*}
    \begin{center}
        \begin{minipage}{1\textwidth}

                \begin{center}
                    \leavevmode \epsfxsize=17cm \epsfbox{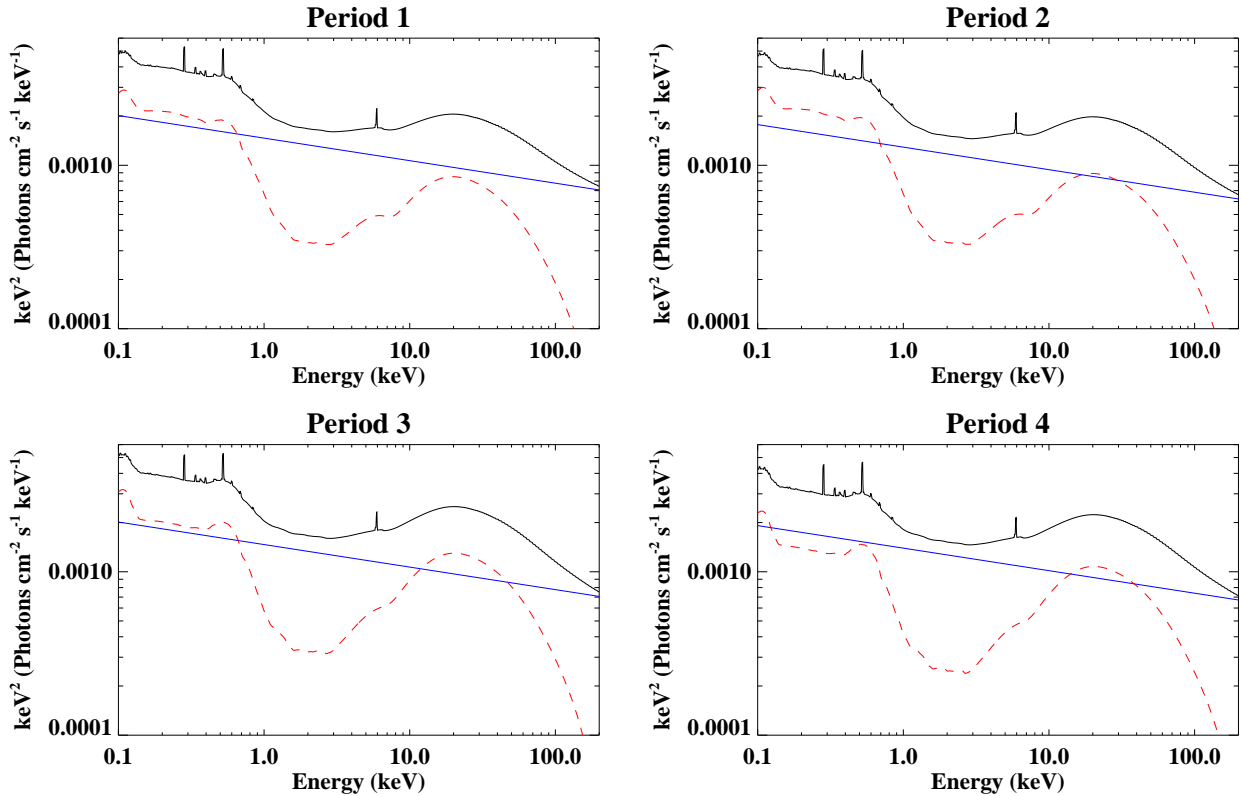}
                \end{center}
            \caption{Each of the set of figures shows the decomposed
            components of model B in a period. The black line above
            shows the overall model without Galactic absorption and
            the edge. The blue line represents a $\Gamma=2.11$ powerlaw.
            The red dash line shows the ionised reflection component.
            The neutral reflection component stays constant throughout the periods and is omitted for the purpose of clarity. }
            \label{eemo_evo}
        \end{minipage}
    \end{center}
\end{figure*}

Based on our time-resolved analysis, we find that Model A, which
still tends to give a high iron abundance, yields a better
$\chi^{2}$ in time-resolved spectra than Model B. This may be due to
the lack of \emph{Swift} spectra that offer a low energy coverage
down to 0.3 keV. The resulting parameters of the time-resolved
spectra are very similar to those obtained in the time-averaged
spectra. The best-fitting values are not exactly the same but all
agree within the 90\% confidence level. The FI XIS and the
\emph{Swift} XRT spectra are not included in fitting the
time-resolved spectra, so slightly different best-fitting results of
parameters are expected.

\subsubsection{Flux Variation} \label{flux_var}

The top two panels of Fig. \ref{flux_evo} show the variation in the
0.01-1000 keV flux level of the powerlaw and the ionised reflection
components, and the bottom panel shows the variation in ionisation
parameter of the reflection component. The ionisation parameter is
defined as $\xi = L_{ion}/nR^{2}$, where $L_{ion}$ is the ionising
luminosity, $n$ is the hydrogen number density, and $R$ is the
distance to the ionising source. $\xi$ is proportional to the
ionising luminosity, which in this case is coming from the powerlaw
continuum. The flux of the powerlaw component varies in time in a
similar manner to the overall  light curve (see Fig. \ref{lc_time})
and is the dominant cause of flux variability in the source. This is
often the case observed in Seyfert 1 galaxies (e.g. MCG-6-30-15).
The ionised reflection flux, on the other hand, remains at a
constant level within the uncertainties through the four periods.
However,  the evolution of ionisation parameter does not follow the
trend laid out by the powerlaw component as expected from the $\xi =
L_{ion}/nR^{2}$ definition of the ionisation parameter if the
product $nR^{2}$ remains constant. Irrespective of which model is
used, i.e. with or without the extra edge, the ionisation parameter
decreases with time.
%\textbf{YOU NEED TO
%DEFINE THE ENERGY RANGE FOR WHICH YOU FOUND THE FLUXES IN FIGURE 9.
%IS IT 0.5 10? OR 0.001 TO 1000KEV..? HOPEFULLY THE SECOND ONE.}
%\textit{It was 0.5 to 200 keV. Now I changed it to 0.01 to 1000 keV and
%re-produce Fig.9. By the way mcg6 is not a NLS1 (very close but not),
%if you want to mention NLS1s here maybe we change the example to be
%Mrk766?}

We depict the evolution of of the powerlaw and the reflection
components in Fig. \ref{eemo_evo}. As the neutral reflection
component is not expected to  vary in these time scales, we only
plot the unabsorbed reflection (ionised plus neutral) component
together with the power law continuum.  From period 1 to period 2,
the reflection component stays more or less constant but the
powerlaw component drops. It can be seen that the drop in the
powerlaw between periods 1 and 2 could, by itself, give rise to an
apparent soft excess in the difference spectrum. The drop in the
ionisation parameter in periods 3 and 4 causes the low energy part
of the reflection to appear lower in comparison to the continuum.
This would of course have the effect of producing an even stronger
soft excess in the difference spectrum.  It is clear from this
qualitative description of the evolution of the reflection component
that the soft excess seen in both the real and difference spectrum
(as well as the broad iron line seen in Fig. \ref{zoomiron}) could
indeed be caused solely by the changes in the ionised reflector and
powerlaw continuum.

We try to fit the \emph{Suzaku} difference spectra presented in Fig.
\ref{pow} with an absorbed powerlaw and an ionised reflection
component. The parameters of the reflection component were set to be
the same as those obtained for the time-averaged spectral analysis.
We find that the simple model indeed fits the difference spectra
well, indicating that the excess shown in the soft difference
spectra could alternatively be caused by the change in the
ionisation parameter. Interestingly, a similar conclusion was
reached by \citet{Reis12}, where the authors find that  changes in
the ionisation parameter was necessary to explain the hard excess
seen in the difference spectra of the Seyfert 1 galaxy NGC 3783. In that
case, the hard excess were due to dramatic changes in the level of
the Compton hump.

\section{Discussion}

\subsection{High Iron Abundance or Absorption Edge}

The iron abundance in a galaxy can be super-solar if there is a high
level of star formation. Some Seyfert I galaxies have super-solar
abundances, for instance, 1H0707-495, which strongly requires
abundances over eight times of solar value \citep{Fabian09}.
However, 1H0707-495 shows the distinct presence of a strong Fe-K
($\sim$ 970 eV) emission line  as well as Fe-L emission
\citep{Fabian09,Zoghbi10,Zoghbi11}. Such strong Fe-K emission
requires a high iron abundance to explain. The high iron abundance
is also the reason why Fe-L emission is seen in the spectrum of
1H0707-495. Nevertheless, the Fe-K line in CBS 126 is not
particularly strong, as would be expected if the abundance was
indeed $>9$. The Fe-L feature is not present in the soft X-ray
spectrum either. Therefore we find it highly unlikely that the
source has such high iron abundances. In addition, it is difficult
to explain the excess above the powerlaw continuum discovered at
low-energies in the difference spectra using model A, as the RMS
spectrum suggest that there are no components contributing extra
variability in the low-energy band.

\begin{figure}
\leavevmode \epsfxsize=8.5cm \epsfbox{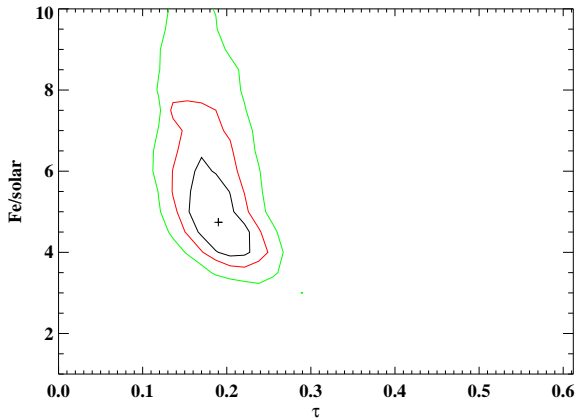} \caption{The
figure shows the contour plot of the depth of the absorption edge
against the iron abundance. } \label{con}
\end{figure}

Fig. \ref{con} shows the contour plot of iron abundance against the
depth of the absorption edge, and contours are plotted at 67\%, 90\%
and 99\% levels. The value of iron abundance spans a wide range from
3 to 10 in the contour plot, implying CBS 126 is likely super-solar,
but the value of iron abundance is not constrained by the current
dataset. We cannot make a correct measurement of the iron abundance
yet. However, the Fe-K line presented in the spectrum is not
particularly strong, the assumption of solar abundance used in Model
B seems to be more reasonable.

%Although the time-resolved spectra tend to be fitted with a high
%iron abundance, solar abundance plus an absorption edge is a more
%self-consistent explanation of the spectra. The low-energy data may
%be important in determining whether the source is super-solar or
%not. The \emph{Swift} XRT spectrum is not of long exposure time, but
%the low-energy spectra are statistically stronger than the
%high-energy spectra. A few data points below 0.5 keV might change
%the model that the data favour. To test this we take the
%\emph{Swift} XRT spectrum out in the time-averaged spectra, and we
%find the \emph{Swift} data does play an important role. If we do not
%include \emph{Swift} data, the dataset has no preference of the
%models (reduced $\chi^{2}$ = 1592/1433 for model A, and 1598/1432
%for model B). This implies that high-quality low-energy data may help
%determine the presence of absorption edge.

\subsection{Possibility of the Existence of Warm Absorbers}

\begin{figure}
\leavevmode \epsfxsize=8.5cm \epsfbox{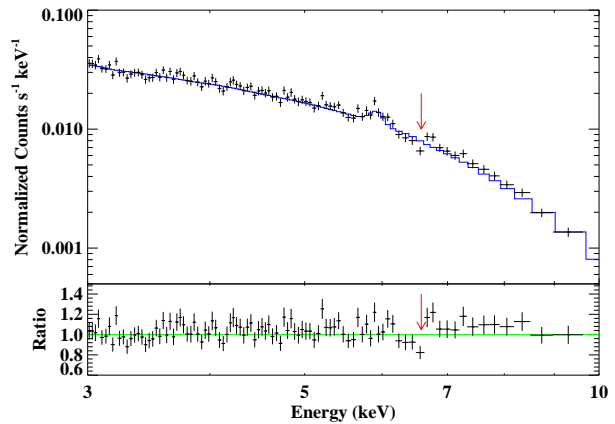} \caption{The figure
shows the \emph{Suzaku} FI XIS spectrum fitted with model B. The red
arrow at $\sim$ 6.6 keV in the figure indicates a possible Fe
{\sevensize XXVI} absorption line. } \label{line}
\end{figure}

Both the time-averaged and the time-resolved spectra can be fitted
successfully by model B. The absorption edge has an energy of $\sim$
0.88 keV, close to the O{\sevensize VIII} edge and Ne {\sevensize
I}, Ne {\sevensize II} edges \citep{Daltabuit72}. The edge is
possibly part of the absorption caused by a warm absorber. Moreover,
it seems there is a weak absorption line at $\sim$ 6.6 keV (see Fig.
\ref{line}) with an EW = $-54^{+44}_{-34}$ eV, which may be a Fe
{\sevensize XXVI} absorption line caused by a highly ionised warm
absorber. This kind of warm absorber produces few absorption lines
in the low-energy spectra. Therefore the absorption edge at $\sim$
0.88 keV, if real, is likely caused by a different warm absorber.
However, the \emph{Suzaku} XIS does not have enough energy
resolution to confirm the existence of warm absorbers. The
\emph{Swift} XRT covers soft X-ray energies down to 0.3 keV but does
not have enough energy resolution either. A more powerful instrument
and an observation with a longer exposure time may be needed to
confirm the assumption. Such an observation will also help index the
absorption edge.

\subsection{Variation of the Ionisation Parameter}

CBS 126 shows enormous variability in long-term X-ray observations
(see Fig. \ref{lc_swift}), and it is also variable in short-term
observations like the one discussed in this paper. The time-resolved
analysis indicates the ionisation parameter varies with time. The
powerlaw component does not vary a lot during the observation, but
the ionisation parameter drops significantly in period 3, and it is
likely that an increase in the hydrogen number density of the disc
or the height of the corona caused the drop in $\xi =
L_{ion}/nR^{2}$. The hydrogen number density of the disc is expected
not to change significantly within a short period of time. The RMS spectrum
indicates that the disc component does not contribute significant
variability. It is more likely that the properties of the accretion
disc remain constant. Assuming $n$ remains constant, $R$ needs to
increase by a factor of $\sim$ 1.4 between period 2 and 3 in order
to explain the drop in ionisation parameter seen here.

As stated in section \ref{flux_var}, different ionisation
parameters produce different amount of soft excess. The excess at
the soft X-ray band seen in the \emph{Suzaku} difference spectra
could be a result of changing ionisation parameter. However,
this again needs to be confirmed by a more detailed observation.

\section{Conclusion}

We test the relativistic reflection model on the new X-ray
observation carried out in late 2010 of the BLS1 galaxy CBS 126. The
spectra show an enormous soft excess in low-energy band, which is
difficult to be explained by only the relativistic reflection
continuum. An iron abundance larger than 9 times the solar value is
needed to produce the soft excess. Based on the information offered
by the difference spectra and the RMS spectrum, a multiplicative
component such as an absorption edge or a warm absorber is more
likely. We find the model composed of a relativistic reflection
continuum with solar abundance and an absorption edge (Model B) is preferred.
To investigate this further, we carried out a
time-resolved analysis to confirm the presence of this feature.
Model B which contains an absorption edge and solar abundance fits
the time-resolved spectra well.
In time-resolved analysis we find the ionisation
parameter drops in period 3, and this is probably caused by an
increase of the height of the corona. The change in ionisation
parameter might also cause the excess at the low-energy band seen in
the difference spectra.

This work shows that the reflection model can be
used in interpreting the spectra of a BLS1 galaxy. We note that a strong soft
excess does not necessarily imply a strong iron K line in the reflection scenario.
We find the edge
model explains both the time-averaged and the time-resolved spectra
well, and it also interprets what we see in the difference and the
RMS spectra. Nevertheless, due to the low signal-to-noise ratio of the data, whether
there are warm absorbers in the source cannot be confirmed yet. The absorption edge is most
likely an O{\sevensize VIII} edge, but the possibility of neon cannot be
ruled out. \emph{Suzaku} and \emph{Swift} do not have enough
resolving power in the low-energy band. An observation by
\emph{Chandra} HETGS or \emph{XMM-Newton} RGS may offer more details
in the low-energy spectra. Future observation of the higher energy spectra with NuSTAR or ASTRO-H
can test the reflection model.

\section*{Acknowledgements}

RCR is supported by NASA through the Einstein Fellowship Program,
grant number PF1-120087. Swift at PSU is supported by NASA contract
NAS5-00136. DG acknowledges support by NASA contract NNX07AH67G. ST
is supported by NASA Grant NNX10AR32G.

\bibliographystyle{mn2e_uw}
\bibliography{cbs126}

\begin{thebibliography}{38}
\expandafter\ifx\csname natexlab\endcsname\relax\def\natexlab#1{#1}\fi

\bibitem[{{Arnaud}(1996)}]{Arnaud96}
{Arnaud} K.~A., 1996, in Astronomical Society of the Pacific Conference Series,
  Vol. 101, Astronomical Data Analysis Software and Systems V, {G.~H.~Jacoby \&
  J.~Barnes}, ed., p.~17

\bibitem[{{Boller} {et~al.}(1996){Boller}, {Brandt}, \& {Fink}}]{Boller96}
{Boller} T., {Brandt} W.~N., {Fink} H., 1996, \aap, 305, 53

\bibitem[{{Brandt} {et~al.}(1997){Brandt}, {Mathur}, \& {Elvis}}]{Brandt97}
{Brandt} W.~N., {Mathur} S., {Elvis} M., 1997, \mnras, 285, L25

\bibitem[{{Burrows} {et~al.}(2005){Burrows}, {Hill}, {Nousek}, {Kennea},
  {Wells}, {Osborne}, {Abbey}, {Beardmore}, {Mukerjee}, {Short}, {Chincarini},
  {Campana}, {Citterio}, {Moretti}, {Pagani}, {Tagliaferri}, {Giommi},
  {Capalbi}, {Tamburelli}, {Angelini}, {Cusumano}, {Br{\"a}uninger}, {Burkert},
  \& {Hartner}}]{Burrows05}
{Burrows} D.~N., {Hill} J.~E., {Nousek} J.~A., {Kennea} J.~A., {Wells} A.,
  {Osborne} J.~P., {Abbey} A.~F., {Beardmore} A., {Mukerjee} K., {Short}
  A.~D.~T., {Chincarini} G., {Campana} S., {Citterio} O., {Moretti} A.,
  {Pagani} C., {Tagliaferri} G., {Giommi} P., {Capalbi} M., {Tamburelli} F.,
  {Angelini} L., {Cusumano} G., {Br{\"a}uninger} H.~W., {Burkert} W., {Hartner}
  G.~D., 2005, \ssr, 120, 165

\bibitem[{{Cardelli} {et~al.}(1989){Cardelli}, {Clayton}, \&
  {Mathis}}]{Cardelli89}
{Cardelli} J.~A., {Clayton} G.~C., {Mathis} J.~S., 1989, \apj, 345, 245

\bibitem[{{Crummy} {et~al.}(2006){Crummy}, {Fabian}, {Gallo}, \&
  {Ross}}]{Crummy06}
{Crummy} J., {Fabian} A.~C., {Gallo} L., {Ross} R.~R., 2006, \mnras, 365, 1067

\bibitem[{{Daltabuit} \& {Cox}(1972)}]{Daltabuit72}
{Daltabuit} E., {Cox} D.~P., 1972, \apj, 177, 855

\bibitem[{{Done} {et~al.}(2012){Done}, {Davis}, {Jin}, {Blaes}, \&
  {Ward}}]{Done12}
{Done} C., {Davis} S.~W., {Jin} C., {Blaes} O., {Ward} M., 2012, \mnras, 420,
  1848

\bibitem[{{Edelson} {et~al.}(2002){Edelson}, {Turner}, {Pounds}, {Vaughan},
  {Markowitz}, {Marshall}, {Dobbie}, \& {Warwick}}]{Edelson02}
{Edelson} R., {Turner} T.~J., {Pounds} K., {Vaughan} S., {Markowitz} A.,
  {Marshall} H., {Dobbie} P., {Warwick} R., 2002, \apj, 568, 610

\bibitem[{{Fabian} {et~al.}(2009){Fabian}, {Zoghbi}, {Ross}, {Uttley}, {Gallo},
  {Brandt}, {Blustin}, {Boller}, {Caballero-Garcia}, {Larsson}, {Miller},
  {Miniutti}, {Ponti}, {Reis}, {Reynolds}, {Tanaka}, \& {Young}}]{Fabian09}
{Fabian} A.~C., {Zoghbi} A., {Ross} R.~R., {Uttley} P., {Gallo} L.~C., {Brandt}
  W.~N., {Blustin} A.~J., {Boller} T., {Caballero-Garcia} M.~D., {Larsson} J.,
  {Miller} J.~M., {Miniutti} G., {Ponti} G., {Reis} R.~C., {Reynolds} C.~S.,
  {Tanaka} Y., {Young} A.~J., 2009, \nat, 459, 540

\bibitem[{{Gehrels} {et~al.}(2004){Gehrels}, {Chincarini}, {Giommi}, {Mason},
  {Nousek}, {Wells}, {White}, {Barthelmy}, {Burrows}, {Cominsky}, {Hurley},
  {Marshall}, {M{\'e}sz{\'a}ros}, {Roming}, {Angelini}, {Barbier}, {Belloni},
  {Campana}, {Caraveo}, {Chester}, {Citterio}, {Cline}, {Cropper}, {Cummings},
  {Dean}, {Feigelson}, {Fenimore}, {Frail}, {Fruchter}, {Garmire}, {Gendreau},
  {Ghisellini}, {Greiner}, {Hill}, {Hunsberger}, {Krimm}, {Kulkarni}, {Kumar},
  {Lebrun}, {Lloyd-Ronning}, {Markwardt}, {Mattson}, {Mushotzky}, {Norris},
  {Osborne}, {Paczynski}, {Palmer}, {Park}, {Parsons}, {Paul}, {Rees},
  {Reynolds}, {Rhoads}, {Sasseen}, {Schaefer}, {Short}, {Smale}, {Smith},
  {Stella}, {Tagliaferri}, {Takahashi}, {Tashiro}, {Townsley}, {Tueller},
  {Turner}, {Vietri}, {Voges}, {Ward}, {Willingale}, {Zerbi}, \&
  {Zhang}}]{Gehrels04}
{Gehrels} N., {Chincarini} G., {Giommi} P., {Mason} K.~O., {Nousek} J.~A.,
  {Wells} A.~A., {White} N.~E., {Barthelmy} S.~D., {Burrows} D.~N., {Cominsky}
  L.~R., {Hurley} K.~C., {Marshall} F.~E., {M{\'e}sz{\'a}ros} P., {Roming}
  P.~W.~A., {Angelini} L., {Barbier} L.~M., {Belloni} T., {Campana} S.,
  {Caraveo} P.~A., {Chester} M.~M., {Citterio} O., {Cline} T.~L., {Cropper}
  M.~S., {Cummings} J.~R., {Dean} A.~J., {Feigelson} E.~D., {Fenimore} E.~E.,
  {Frail} D.~A., {Fruchter} A.~S., {Garmire} G.~P., {Gendreau} K., {Ghisellini}
  G., {Greiner} J., {Hill} J.~E., {Hunsberger} S.~D., {Krimm} H.~A., {Kulkarni}
  S.~R., {Kumar} P., {Lebrun} F., {Lloyd-Ronning} N.~M., {Markwardt} C.~B.,
  {Mattson} B.~J., {Mushotzky} R.~F., {Norris} J.~P., {Osborne} J., {Paczynski}
  B., {Palmer} D.~M., {Park} H.-S., {Parsons} A.~M., {Paul} J., {Rees} M.~J.,
  {Reynolds} C.~S., {Rhoads} J.~E., {Sasseen} T.~P., {Schaefer} B.~E., {Short}
  A.~T., {Smale} A.~P., {Smith} I.~A., {Stella} L., {Tagliaferri} G.,
  {Takahashi} T., {Tashiro} M., {Townsley} L.~K., {Tueller} J., {Turner}
  M.~J.~L., {Vietri} M., {Voges} W., {Ward} M.~J., {Willingale} R., {Zerbi}
  F.~M., {Zhang} W.~W., 2004, \apj, 611, 1005

\bibitem[{{Gierli{\'n}ski} \& {Done}(2004)}]{Gierlinski04}
{Gierli{\'n}ski} M., {Done} C., 2004, \mnras, 349, L7

\bibitem[{{Gierli{\'n}ski} \& {Done}(2006)}]{Gierlinski06}
---, 2006, \mnras, 371, L16

\bibitem[{{Godet} {et~al.}(2009){Godet}, {Beardmore}, {Abbey}, {Osborne},
  {Cusumano}, {Pagani}, {Capalbi}, {Perri}, {Page}, {Burrows}, {Campana},
  {Hill}, {Kennea}, \& {Moretti}}]{Godet09}
{Godet} O., {Beardmore} A.~P., {Abbey} A.~F., {Osborne} J.~P., {Cusumano} G.,
  {Pagani} C., {Capalbi} M., {Perri} M., {Page} K.~L., {Burrows} D.~N.,
  {Campana} S., {Hill} J.~E., {Kennea} J.~A., {Moretti} A., 2009, \aap, 494,
  775

\bibitem[{{Grupe} {et~al.}(1998{\natexlab{a}}){Grupe}, {Beuermann}, {Thomas},
  {Mannheim}, \& {Fink}}]{Grupe98}
{Grupe} D., {Beuermann} K., {Thomas} H.-C., {Mannheim} K., {Fink} H.~H.,
  1998{\natexlab{a}}, \aap, 330, 25

\bibitem[{{Grupe} {et~al.}(2010){Grupe}, {Komossa}, {Leighly}, \&
  {Page}}]{Grupe10}
{Grupe} D., {Komossa} S., {Leighly} K.~M., {Page} K.~L., 2010, \apjs, 187, 64

\bibitem[{{Grupe} {et~al.}(2001){Grupe}, {Thomas}, \& {Beuermann}}]{Grupe01}
{Grupe} D., {Thomas} H.-C., {Beuermann} K., 2001, \aap, 367, 470

\bibitem[{{Grupe} {et~al.}(2004){Grupe}, {Wills}, {Leighly}, \&
  {Meusinger}}]{Grupe04}
{Grupe} D., {Wills} B.~J., {Leighly} K.~M., {Meusinger} H., 2004, \aj, 127, 156

\bibitem[{{Grupe} {et~al.}(1998{\natexlab{b}}){Grupe}, {Wills}, {Wills}, \&
  {Beuermann}}]{Grupe98b}
{Grupe} D., {Wills} B.~J., {Wills} D., {Beuermann} K., 1998{\natexlab{b}},
  \aap, 333, 827

\bibitem[{{Hill} {et~al.}(2004){Hill}, {Burrows}, {Nousek}, {Abbey}, {Ambrosi},
  {Br{\"a}uninger}, {Burkert}, {Campana}, {Cheruvu}, {Cusumano}, {Freyberg},
  {Hartner}, {Klar}, {Mangels}, {Moretti}, {Mori}, {Morris}, {Short},
  {Tagliaferri}, {Watson}, {Wood}, \& {Wells}}]{Hill04}
{Hill} J.~E., {Burrows} D.~N., {Nousek} J.~A., {Abbey} A.~F., {Ambrosi} R.~M.,
  {Br{\"a}uninger} H.~W., {Burkert} W., {Campana} S., {Cheruvu} C., {Cusumano}
  G., {Freyberg} M.~J., {Hartner} G.~D., {Klar} R., {Mangels} C., {Moretti} A.,
  {Mori} K., {Morris} D.~C., {Short} A.~D.~T., {Tagliaferri} G., {Watson}
  D.~J., {Wood} P., {Wells} A.~A., 2004, in Society of Photo-Optical
  Instrumentation Engineers (SPIE) Conference Series, Vol. 5165, Society of
  Photo-Optical Instrumentation Engineers (SPIE) Conference Series,
  {K.~A.~Flanagan \& O.~H.~W.~Siegmund}, ed., pp. 217--231

\bibitem[{{Jin} {et~al.}(2012){Jin}, {Ward}, {Done}, \& {Gelbord}}]{Jin12}
{Jin} C., {Ward} M., {Done} C., {Gelbord} J., 2012, \mnras, 420, 1825

\bibitem[{{Miniutti} \& {Fabian}(2004)}]{Miniutti04}
{Miniutti} G., {Fabian} A.~C., 2004, \mnras, 349, 1435

\bibitem[{{Mitsuda} {et~al.}(1984){Mitsuda}, {Inoue}, {Koyama}, {Makishima},
  {Matsuoka}, {Ogawara}, {Suzuki}, {Tanaka}, {Shibazaki}, \& {Hirano}}]{diskbb}
{Mitsuda} K., {Inoue} H., {Koyama} K., {Makishima} K., {Matsuoka} M., {Ogawara}
  Y., {Suzuki} K., {Tanaka} Y., {Shibazaki} N., {Hirano} T., 1984, \pasj, 36,
  741

\bibitem[{{Poole} {et~al.}(2008){Poole}, {Breeveld}, {Page}, {Landsman},
  {Holland}, {Roming}, {Kuin}, {Brown}, {Gronwall}, {Hunsberger}, {Koch},
  {Mason}, {Schady}, {vanden Berk}, {Blustin}, {Boyd}, {Broos}, {Carter},
  {Chester}, {Cucchiara}, {Hancock}, {Huckle}, {Immler}, {Ivanushkina},
  {Kennedy}, {Marshall}, {Morgan}, {Pandey}, {de Pasquale}, {Smith}, \&
  {Still}}]{Poole08}
{Poole} T.~S., {Breeveld} A.~A., {Page} M.~J., {Landsman} W., {Holland} S.~T.,
  {Roming} P., {Kuin} N.~P.~M., {Brown} P.~J., {Gronwall} C., {Hunsberger} S.,
  {Koch} S., {Mason} K.~O., {Schady} P., {vanden Berk} D., {Blustin} A.~J.,
  {Boyd} P., {Broos} P., {Carter} M., {Chester} M.~M., {Cucchiara} A.,
  {Hancock} B., {Huckle} H., {Immler} S., {Ivanushkina} M., {Kennedy} T.,
  {Marshall} F., {Morgan} A., {Pandey} S.~B., {de Pasquale} M., {Smith} P.~J.,
  {Still} M., 2008, \mnras, 383, 627

\bibitem[{{Pounds} {et~al.}(1995){Pounds}, {Done}, \& {Osborne}}]{Pounds95}
{Pounds} K.~A., {Done} C., {Osborne} J.~P., 1995, \mnras, 277, L5

\bibitem[{{Puchnarewicz} {et~al.}(1992){Puchnarewicz}, {Mason}, {Cordova},
  {Kartje}, {Brabduardi}, {Puchnarewicz}, {Mason}, {Cordova}, {Kartje},
  {Branduardi-Raymont}, {Mittaz}, {Murdin}, \& {Allington-Smith}}]{Puch92}
{Puchnarewicz} E.~M., {Mason} K.~O., {Cordova} F.~A., {Kartje} J., {Brabduardi}
  A.~A., {Puchnarewicz} E.~M., {Mason} K.~O., {Cordova} F.~A., {Kartje} J.,
  {Branduardi-Raymont} G., {Mittaz} J.~P.~D., {Murdin} P.~G., {Allington-Smith}
  J., 1992, \mnras, 256, 589

\bibitem[{{Reis} {et~al.}(2012){Reis}, {Fabian}, {Reynolds}, {Brenneman},
  {Walton}, {Trippe}, {Miller}, {Mushotzky}, \& {Nowak}}]{Reis12}
{Reis} R.~C., {Fabian} A.~C., {Reynolds} C.~S., {Brenneman} L.~W., {Walton}
  D.~J., {Trippe} M., {Miller} J.~M., {Mushotzky} R.~F., {Nowak} M.~A., 2012,
  \apj, 745, 93

\bibitem[{{Roming} {et~al.}(2005){Roming}, {Kennedy}, {Mason}, {Nousek}, {Ahr},
  {Bingham}, {Broos}, {Carter}, {Hancock}, {Huckle}, {Hunsberger}, {Kawakami},
  {Killough}, {Koch}, {McLelland}, {Smith}, {Smith}, {Soto}, {Boyd},
  {Breeveld}, {Holland}, {Ivanushkina}, {Pryzby}, {Still}, \&
  {Stock}}]{Roming05}
{Roming} P.~W.~A., {Kennedy} T.~E., {Mason} K.~O., {Nousek} J.~A., {Ahr} L.,
  {Bingham} R.~E., {Broos} P.~S., {Carter} M.~J., {Hancock} B.~K., {Huckle}
  H.~E., {Hunsberger} S.~D., {Kawakami} H., {Killough} R., {Koch} T.~S.,
  {McLelland} M.~K., {Smith} K., {Smith} P.~J., {Soto} J.~C., {Boyd} P.~T.,
  {Breeveld} A.~A., {Holland} S.~T., {Ivanushkina} M., {Pryzby} M.~S., {Still}
  M.~D., {Stock} J., 2005, \ssr, 120, 95

\bibitem[{{Roming} {et~al.}(2009){Roming}, {Koch}, {Oates}, {Porterfield},
  {Vanden Berk}, {Boyd}, {Holland}, {Hoversten}, {Immler}, {Marshall}, {Page},
  {Racusin}, {Schneider}, {Breeveld}, {Brown}, {Chester}, {Cucchiara},
  {DePasquale}, {Gronwall}, {Hunsberger}, {Kuin}, {Landsman}, {Schady}, \&
  {Still}}]{Roming09}
{Roming} P.~W.~A., {Koch} T.~S., {Oates} S.~R., {Porterfield} B.~L., {Vanden
  Berk} D.~E., {Boyd} P.~T., {Holland} S.~T., {Hoversten} E.~A., {Immler} S.,
  {Marshall} F.~E., {Page} M.~J., {Racusin} J.~L., {Schneider} D.~P.,
  {Breeveld} A.~A., {Brown} P.~J., {Chester} M.~M., {Cucchiara} A.,
  {DePasquale} M., {Gronwall} C., {Hunsberger} S.~D., {Kuin} N.~P.~M.,
  {Landsman} W.~B., {Schady} P., {Still} M., 2009, \apj, 690, 163

\bibitem[{{Ross} \& {Fabian}(2005)}]{Ross05}
{Ross} R.~R., {Fabian} A.~C., 2005, \mnras, 358, 211

\bibitem[{{Schlegel} {et~al.}(1998){Schlegel}, {Finkbeiner}, \&
  {Davis}}]{Schlegel98}
{Schlegel} D.~J., {Finkbeiner} D.~P., {Davis} M., 1998, \apj, 500, 525

\bibitem[{{Schurch} \& {Done}(2007)}]{Schurch07}
{Schurch} N.~J., {Done} C., 2007, \mnras, 381, 1413

\bibitem[{{Shen} {et~al.}(2011){Shen}, {Richards}, {Strauss}, {Hall},
  {Schneider}, {Snedden}, {Bizyaev}, {Brewington}, {Malanushenko},
  {Malanushenko}, {Oravetz}, {Pan}, \& {Simmons}}]{Shen11}
{Shen} Y., {Richards} G.~T., {Strauss} M.~A., {Hall} P.~B., {Schneider} D.~P.,
  {Snedden} S., {Bizyaev} D., {Brewington} H., {Malanushenko} V.,
  {Malanushenko} E., {Oravetz} D., {Pan} K., {Simmons} A., 2011, \apjs, 194, 45

\bibitem[{{Titarchuk}(1994)}]{comptt}
{Titarchuk} L., 1994, \apj, 434, 570

\bibitem[{{Voges} {et~al.}(1999){Voges}, {Aschenbach}, {Boller},
  {Br{\"a}uninger}, {Briel}, {Burkert}, {Dennerl}, {Englhauser}, {Gruber},
  {Haberl}, {Hartner}, {Hasinger}, {K{\"u}rster}, {Pfeffermann}, {Pietsch},
  {Predehl}, {Rosso}, {Schmitt}, {Tr{\"u}mper}, \& {Zimmermann}}]{Voges99}
{Voges} W., {Aschenbach} B., {Boller} T., {Br{\"a}uninger} H., {Briel} U.,
  {Burkert} W., {Dennerl} K., {Englhauser} J., {Gruber} R., {Haberl} F.,
  {Hartner} G., {Hasinger} G., {K{\"u}rster} M., {Pfeffermann} E., {Pietsch}
  W., {Predehl} P., {Rosso} C., {Schmitt} J.~H.~M.~M., {Tr{\"u}mper} J.,
  {Zimmermann} H.~U., 1999, \aap, 349, 389

\bibitem[{{Wilms} {et~al.}(2000){Wilms}, {Allen}, \& {McCray}}]{Wilms00}
{Wilms} J., {Allen} A., {McCray} R., 2000, \apj, 542, 914

\bibitem[{{Zoghbi} {et~al.}(2010){Zoghbi}, {Fabian}, {Uttley}, {Miniutti},
  {Gallo}, {Reynolds}, {Miller}, \& {Ponti}}]{Zoghbi10}
{Zoghbi} A., {Fabian} A.~C., {Uttley} P., {Miniutti} G., {Gallo} L.~C.,
  {Reynolds} C.~S., {Miller} J.~M., {Ponti} G., 2010, \mnras, 401, 2419

\bibitem[{{Zoghbi} {et~al.}(2011){Zoghbi}, {Uttley}, \& {Fabian}}]{Zoghbi11}
{Zoghbi} A., {Uttley} P., {Fabian} A.~C., 2011, \mnras, 412, 59

\end{thebibliography}

\label{lastpage}
\end{document}